\documentclass[twocolumn,showpacs,preprintnumbers,amsmath,amssymb]{revtex4}

\usepackage{graphicx}
\usepackage{dcolumn}
\usepackage{bm}

\begin{document}
\title{Efficient routing on complex networks}
\author{Gang Yan$^{1}$}
\author{Tao Zhou$^{1,2}$}
\email{zhutou@ustc.edu}
\author{Bo Hu$^{2}$}
\author{Zhong-Qian Fu$^{1}$}
\author{Bing-Hong Wang$^{2}$}
\affiliation{%
$^{1}$Department of Electronic Science and Technology,\\
$^{2}$Nonlinear Science Center and Department of Modern Physics,\\
University of Science and Technology of China, \\
Hefei Anhui, 230026, PR China \\
}%
\date{\today}

\begin{abstract}
In this letter, we propose a new routing strategy to improve the
transportation efficiency on complex networks. Instead of using
the routing strategy for shortest path, we give a generalized
routing algorithm to find the so-called {\it efficient path},
which considers the possible congestion in the nodes along actual
paths. Since the nodes with largest degree are very susceptible to
traffic congestion, an effective way to improve traffic and
control congestion, as our new strategy, can be as redistributing
traffic load in central nodes to other non-central nodes.
Simulation results indicate that the network capability in
processing traffic is improved more than 10 times by optimizing
the efficient path, which is in good agreement with the analysis.
\end{abstract}


\maketitle
 Since the seminal work on scale-free networks by
Barab\'asi and Albert (BA model) \cite{BA} and on the small-world
phenomenon by Watts and Strogatz \cite{WS}, the structure and
dynamics on complex networks have recently attracted a tremendous
amount of interest and devotion from physics community. The
increasing importance of large communication networks such as the
Internet, upon which our society survives, calls for the need for
high efficiency in handling and delivering information. In this
light, to find optimal strategies for traffic routing is one of
the important issues we have to address.

There have been many previous studies to understand and control
traffic congestion on networks, with a basic assumption that the
network has a homogeneous structure\cite{IEEE}. However, many
realistic networks like the Internet display both scale-free and
small-world features, and thus it is of great interest to study
the effect of network topology on traffic flow and the effect of
traffic on network evolution. Guimer\'a et al present a formalism
that can cope simultaneously with the searching and traffic
dynamics in parallel transportation systems\cite{PRL2}. This
formalism can be used to optimal networks structure under a local
search algorithm, while to obtain the formalism one should know
the global information of the whole networks. Holme and Kim
provide an in-depth analysis on the vertex/edge overload cascading
breakdowns based on evolving networks, and suggest a method to
avoid such avalanches\cite{Holme}. Since the load of a certain
vertex/edge is defined at its betweenness\cite{Kahng}, there is
also a latent assumption that the routing protocols involved
global knowledge of networks. By using global and dynamical
searching algorithm aiming at shortest paths, Zhao et al provide
the theoretical estimates of the communication capacity
\cite{LYC}. Since the global information is usually unavailable in
large-scale networks, Tadi\'{c} et al investigate the traffic
dynamics on WWW network model\cite{TadicPA} based on local
knowledge, providing insights into the relationship of global
statistical properties and microscopic density
fluctuations\cite{Tadic}. The routing strategies for
Internet\cite{Routing} and disordered networks\cite{Gene} are also
studied.  Another interesting issue is the interplay of traffic
dynamics and network structures, which suggests a new scenario of
network evolutionary\cite{BBV}.
\begin{figure}
\scalebox{0.75}[0.6]{\includegraphics{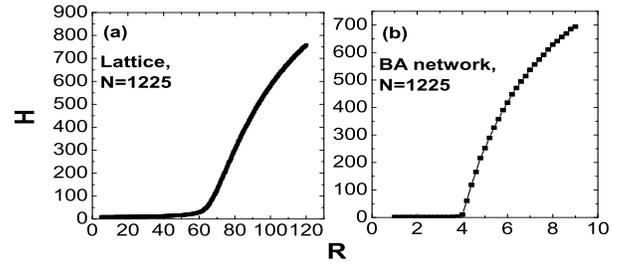}}
\caption{\label{fig:epsart} The order parameter $H$ versus $R$ for
two-dimensional lattice (a) and BA networks (b) with the same size
$N=1225$. The routing algorithm at the shortest path yields
$R_{Lattice}\approx60$ and $R_{BA}\approx4.0$.}
\end{figure}
\begin{figure}
\scalebox{0.80}[0.60]{\includegraphics{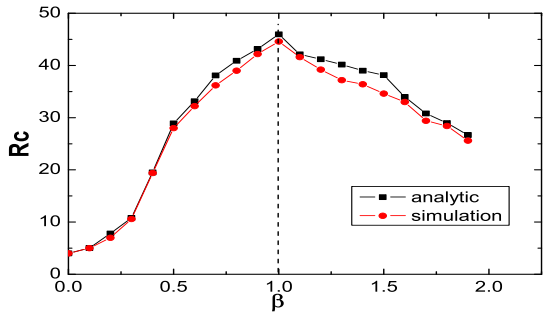}}
\caption{\label{fig:epsart} (Color online). The critical $R_c$
versus $\beta$ for scale-free networks with size $N=1225$. Both
Simulation and analysis indicate that the maximum of $R_c$
corresponds to $\beta\approx1.0$. The data shown here are the
average over 10 independent runs.}
\end{figure}

In this context, the information processor are routers that have
the same function in the postal serves. For simplicity, we treat
all the nodes as both hosts and routers \cite{PRL2}. In
communication networks, routers deliver data packets by ensuring
that all converge to a best estimate of the path leading to each
destination address. In other words, the routing process takes
place following according to the criterion of the shortest
available path from a given source to its destination. When the
network size $N$ is not too large, it is possible to calculate all
the shortest paths between any nodes, and thus the traffic system
can use a fixed routing table to process information. As for any
pair of source and destination, there may be several shortest
paths between them. We can randomly choose one of them and put it
into the fixed routing table which is followed by all the
information packets. Though it becomes impractical in huge
communication systems, the fixed routing algorithm is widely used
in mid- or small-systems \cite{xxx}. It is because that the fixed
routing method has obvious advantages in economical and technical
costs, compared with dynamical routing algorithm and information
feed-back mechanism. The model is described as follows: at each
time step, there are $R$ packets generated in the system, with
randomly chosen sources and destinations. It is assumed that all
the routers have the same capabilities in delivering and handling
information packets, that is, at each time step all the nodes can
deliver at most $C$ packets one step toward their destinations
according to the fixed routing table. We set $C=1$ for simplicity.
A packet, once reaching its destination, is removed from the
system. We are most interested in the critical value $R_c$ of
information generation (as measured by the number of packets
created within the network per unit time) where a phase transition
takes place from free flow to congested traffic. This critical
value can best reflect the maximum capability of a system handling
its traffic. In particular, for $R<R_c$, the numbers of created
and delivered packets are balanced, leading to a steady free
traffic flow. For $R>R_c$, traffic congestion occurs as the number
of accumulated packets increases with time, simply for that the
capacities of nodes for delivering packets are limited. We use the
order parameter to characterize the phase transition
\begin{figure}
\scalebox{0.42}[0.5]{\includegraphics{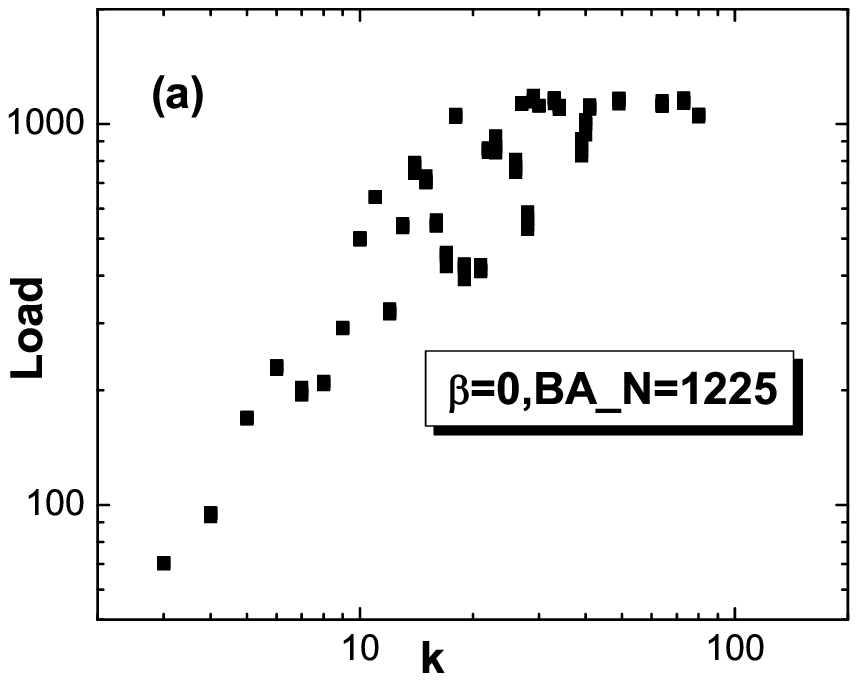}}
\scalebox{0.42}[0.5]{\includegraphics{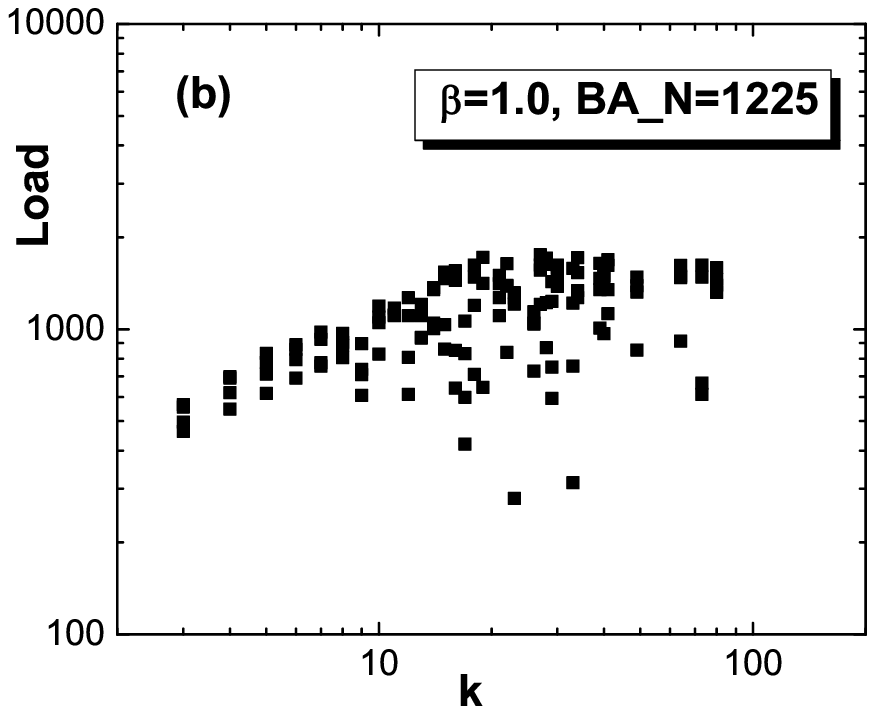}}
\caption{\label{fig:epsart} The load distribution when congestion
occurs for BA network with size $N=1225$. (a) The case of
$\beta=0$ where $R_c=4.0$ and we set $R=10$. (b) The case of
$\beta=1$ where $R_c=45$ and we set $R=60$.}
\end{figure}
\begin{equation}
H(R)=\lim_{t\rightarrow\infty}\frac{C}{R}\frac{\langle\Delta W\rangle}{\Delta t},
\end{equation}
where $\Delta W=W(t+\Delta t)-W(t)$ with $\langle\cdots\rangle$
indicates average over time windows of width $\Delta t$, and W(t)
is the total number of packets in the network at time $t$. Fig. 1
shows the order parameter $H$ versus $R$ for (a) the
two-dimensional lattice with periodical boundary condition and (b)
the scale-free BA network with average degree $\langle k
\rangle=4$ \cite{BA}, given all the packets follows their shortest
paths. We see that the critical point $R_c$ in lattice is much
larger than that in scale-free network. This phenomena can be
simply explained by their different betweenness centralities (BC)
\cite{Kahng}. The BC of a node $v$ is defined as
\begin{equation}
g(v)=\sum_{s\neq t}\frac{\sigma_{st}(v)}{\sigma_{st}},
\end{equation}
where $\sigma_{st}$ is the number of shortest paths
going from $s$ to $t$ and $\sigma_{st}(v)$ is the number of
shortest paths going from $s$ to $t$ and passing through $v$. This
definition means that central nodes are part of more shortest
paths within the network than peripheral nodes. Moreover, BC gives
in transport networks an estimate of the traffic handled by the
vertices, assuming that the number of shortest paths is a zero-th
order approximation to the frequency of use of a given node. It is
generally useful to represent the average BC for vertices of the
same degree
\begin{equation}
g(k)=\frac{1}{N_k}\sum_{v,k_v=k}g(v),
\end{equation}
where $N_k$ denotes the number of nodes with degree $k$. For most
networks, $g(k)$ is strongly correlated with $k$. In general, the
larger the degree, the larger the centrality. For scale-free
networks it has been shown that the centrality scales with $k$ as
$g(k)\sim k^\mu$ where $\mu$ depends on the network. Hence in the
scale-free networks, the betweenness distribution also obeys a
power-law form. In comparison, the BC in lattice will behave a
homogeneous distribution. Noticeably, in scale-free networks,
traffic congestion generally occurs at nodes with the largest
degree (or BC), and immediately spreads over all the nodes. When
all the packets follow their shortest paths, it will easily lead
to the overload of the heavily-linked router, which is just the
key of traffic congestion. To alleviate the congestion, a feasible
and effective way is to bypass such high-degree nodes in traffic
routing design. This leads us to question the commonly used
shortest-path routing mechanism.

\begin{figure}
\scalebox{0.8}[0.6]{\includegraphics{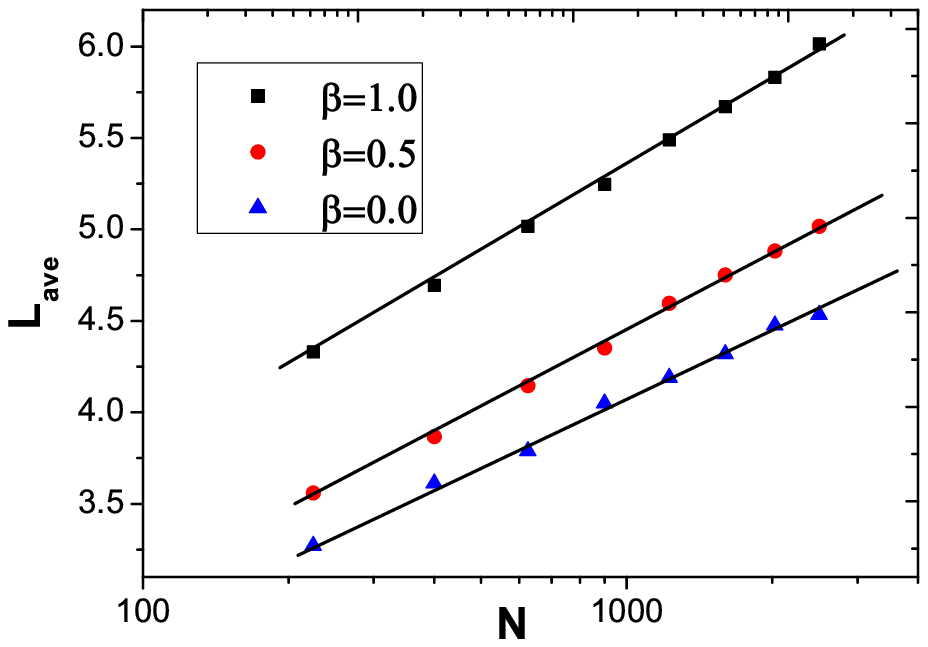}}
\scalebox{0.73}[0.6]{\includegraphics{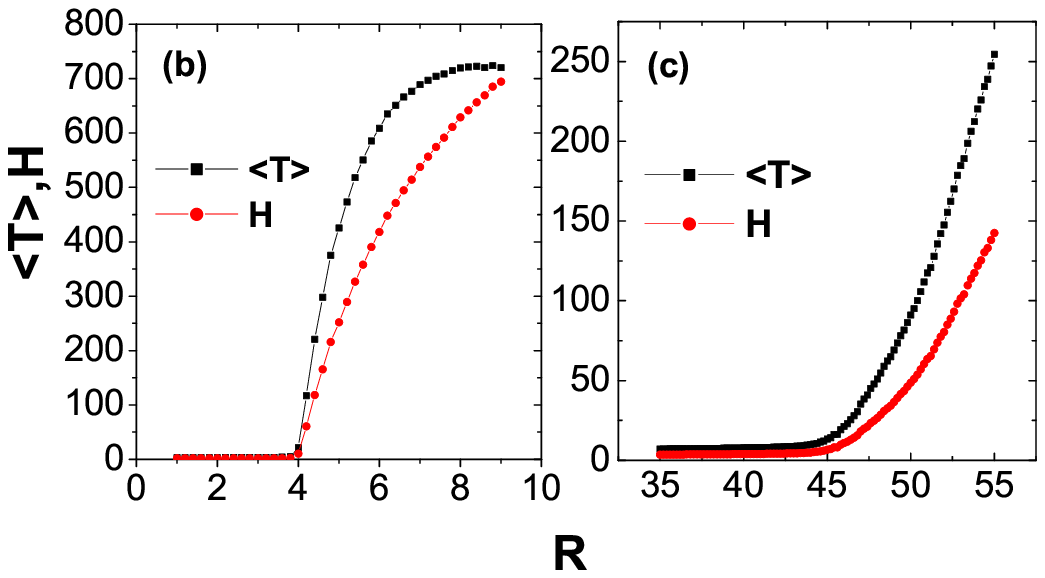}}
\caption{\label{fig:epsart} (Color online). (a) The average actual
path length $L_{ave}$ versus the network size $N$ under various
values of $\beta$, by using the efficient path routing. (b) and
(c) show $\langle T \rangle$ and $H$ versus $R$ for $\beta=0.0$
and $\beta=1.0$ respectively.}
\end{figure}

Actually, the path with shortest length is not necessarily the
quickest way, considering the presence of possible traffic
congestion and waiting time along the shortest path (by
``shortest" we mean the path with smallest number of links).
Obviously, nodes with larger connections are more likely to bear
traffic congestion. Intuitively, a packet will by average spends
more time to travel through a node with higher BC (for waiting
time). All too often, bypassing those high-BC nodes, a packet may
reach its destination quicker than taking the shortest path. In
order to find the optimal routing strategy, we define the
``efficient path". For any path between nodes $i$ \& $j$ as
$P(i\rightarrow j):=i\equiv x_0, x_1, \cdots x_{n-1}, x_n \equiv
j$, we define
\begin{equation}
L(P(i\rightarrow j):\beta)=\sum_{i=0}^{n-1}k(x_i)^{\beta}.
\end{equation}
For any given $\beta$, the efficient path between $i$ and $j$ is
corresponding to the route that makes the sum $L(P(i\rightarrow
j):\beta)$ minimum. Obviously, $L_{min}(\beta=0)$ recovers the
traditionally shortest path length. We expect that the system
behaves better under the routing rule with $\beta>0$ than it does
traditionally, and we aim to find the optimal $\beta$ in this
letter. In the following, the fixed routing table is designed on
the basis of efficient path. If there are several efficient paths
between two nodes, the one is chosen at random. We are now
interested in determining the phase-transition point $R_c$ under
various $\beta$, in order to address which kind of routing
strategy is more flexible to traffic congestion, and therefore
find the optimal $\beta$.

Aiming to estimate the value of $R_c$ for different $\beta$, we define
the efficient betweennes centralities(EBC) of a node $\upsilon$ as,
\begin{equation}
g^{\beta}(\upsilon)=\sum_{s\neq t}\frac{\sigma^{\beta}_{st}(\upsilon)}{\sigma^{\beta}_{st}}
\end{equation}
where $\sigma^{\beta}_{st}$ is the number of efficient paths
for a given $\beta$ going from $s$ to $t$ and $\sigma^{\beta}_{st}(v)$ is the number of
efficient paths for a given $\beta$ going from $s$ to $t$ and passing through $v$.
Since congestion occurs at the node with the largest betweenness, $R_c$ can be estimated as\cite{LYC,PRL2},
\begin{equation}
R_c=\frac{N(N-1)}{g_{max}}
\end{equation}
where $N$ is the size of the network and $g_{max}$ is the largest BC of the network.
Similarly, for different $\beta$, we can estimate the $R_c(\beta)$ as,
\begin{equation}
R_c(\beta)=\frac{N(N-1)}{g^{\beta}_{max}}
\end{equation}
where $g^{\beta}_{max}$ is the largest EBC for a given $\beta$.

In Fig. 2 we report the simulation results for the critical value
$R_c$ as a function of $\beta$ on BA networks, which is in good
agreement with the analysis. As one can see, $R_c$ first increases
with $\beta$ and then decreases, with the maximum of $R_c$
corresponds to $\beta\approx1.0$. In comparison with the shortest
path routing case (i.e. $\beta=0$), the capability of the network
in freely handling information is greatly improved, from
$R_c\approx4.0$ when $\beta=0$ to $R_c\approx45$ when $\beta=1.0$;
more than ten times. This result suggests us the effectiveness of
the routing strategy by our efficient path length. Fig. 3 shows
the optimized behavior of (b) our efficient path routing in load
distribution when congestion just occurs, in comparison with that
of (a) the shortest path routing mechanism. Clearly, the heavy
load on central nodes (with highest connectivity) is strongly
redistributed to those nodes with less degree by using efficient
path routing table. We also report in Fig.4(a) the average actual
path length $L_{ave}$ versus the network size $N$ under various
values of $\beta$. As one can see, although $L_{ave}$ increases
with $\beta$, the small-world property $L_{ave}\sim \ln N$ is
still kept. The system capability in processing information is
considerably enhanced at the cost of increasing the average
routing path length. Such a sacrifice may be worthwhile when a
system requires large $R_c$. Moreover, we investigate the average
transporting time $\langle T \rangle$ of packets. The results in
Fig.4(b) and Fig.4(c) show that $\langle T \rangle$ and $H$
indicate the same critical value $R_c$.

To realize the routing strategy we have studied, each router must
have the complete knowledge of the network topology, which is
often difficult for large-scale systems. Anyway, it is possible to
divide one large system into several autonomous subsystems in
which every router has its local topological knowledge. Thus, the
hierarchical structure of the network will make possible the
implementation of our routing strategy. This paper has mainly
discussed how to effectively design routing algorithm when the
capabilities of processing information are the same for all the
nodes. To account for the network topology, one can assume that
the capabilities for processing information are different for
different nodes, depending on the numbers of links or the number
of the shortest paths passing through them \cite{LYC}. In
addition, the shortest path is shortest just in a topological
sense; in practice, it is not necessarily the best. As for a
single packet, its best routing as we have argued is not
absolutely the shortest path. From the systematic view, the total
information load that a communication network can freely handle
without congestion depends on all the packets can reach their
destinations in a systematically optimal time. We use $R_c$ to
denote the upper limit of the total information load that a
communication system can handle without congestion. This parameter
reflect the system capability in processing information under
certain routing strategy. An effective way found to alleviate
traffic congestion for scale-free networks is to make the heavily
linked nodes as powerful and efficient as possible for processing
information. This is further supported by examining the effect of
enhancing the capabilities of these nodes. In addition, some
previous models aiming at communication networks are closer to
reality than BA networks, such as directed scale-free model for
World-Wide-Web \cite{TadicPA} and positive-feedback preference
model for Internet \cite{Zhou2004}. To investigate the present
traffic model and routing strategy onto these network models are
significant in practice. These will be among the future works.

While our model is based on computer networks, we expect it to be
relevant to other practical networks in general. Our studies may
be useful for designing communication protocols for complex
networks, considering there appears no increase in its algorithmic
complexity. The optimized routing strategy studied in this paper
can be easily implemented in practice.

Many previous works focus on the relationship between the
distribution of BC and the capability of communication networks,
with a latent assumption that the information packets go along the
shortest paths from source to destination. Therefore, the BC is
always considered as a static topological measure of networks.
Here we argue that, this quantity is determined both by the
routing algorithm and network topology, thus one should pay more
attention to the design of routing strategies. We believe this
work may enlighten readers on this subject and be helpful for
understanding the intrinsic mechanism of network traffic. Finally,
it is worthwhile to emphasize that, we have found some evidences
indicating there may exist some common features between network
traffic and synchronization in dynamical level, thus the present
method may be also useful for enhancing the network
synchronizability\cite{Motter2005}.

The authors wish to thank Dr.Wen-Xu Wang for discussions. This
work has been partially supported by the National Natural Science
Foundation of China under Grant No.71471033, 10472116, and
70271070, and the Specialized Research Fund for the Doctoral
Program of Higher Education (SRFDP No.20020358009)

\end{document}